\documentclass[aps,prc,twocolumn,nofootinbib]{revtex4-2}
\usepackage[T1]{fontenc}
\usepackage{times}
\usepackage{amsmath,amssymb}
\usepackage{hyperref}  
\usepackage{graphicx}
\usepackage{booktabs}
\usepackage{mathtools}
\usepackage{subcaption}
\usepackage[dvipsnames]{xcolor}
\usepackage{slashed}
\usepackage{bm}
\usepackage{physics}
\usepackage{hyperref}
\usepackage{caption}
\begin{document}

\title{Suppression of genuine tripartition in heavy nuclei: A self-consistent perspective}

\author{Y. Jaganathen$^{1}$}
\author{J. Skalski$^{1}$}
\email{janusz.skalski@ncbj.gov.pl}  

\affiliation{$^{1}$National Centre for Nuclear Research, Pasteura 7, 02-093 Warsaw, Poland}

\date{\today}  

\begin{abstract}
We investigate the ternary fissions of $^{252}$Cf (spontaneous) and $^{236}$U 
 (neutron-induced) into medium-mass fragments, as reported by the Dubna group,
 using the Hartree-Fock plus BCS method with the SLy6 Skyrme 
 interaction. Compared to microscopic-macroscopic methods used so far, this self-consistent approach provides a greater flexibility of nuclear shapes.
 
Our working hypothesis is that the shape evolution proceeds while the system is
 still mononuclear. The results show that a ternary fission valley emerges for 
 intermediate elongations of the middle fragment, only when its mass is 
 strongly constrained. This ternary mode is dynamically suppressed 
 by the competition with the dominant binary decay channel.  
  This suggests that a description based solely on quantum tunneling through 
  the energy barrier is insufficient to evaluate its probability. 
  
 To quantify this suppression, we apply a simple Langevin-type model of 
 overdamped motion with constant damping and temperature, supplemented by a 
 basic estimate of quantum tunneling where relevant.
 Under assumptions expected to yield an upper bound, we find the  probability 
 of ternary fission per binary decay to be on the order of
   $10^{-8}-10^{-9}$ for $^{252}$Cf and $10^{-10}-10^{-11}$ for $^{236}$U.

\vspace{3pt}

\textbf{Note:} This is the authors' manuscript published in Physics Letters B. \href{https://doi.org/10.1016/j.physletb.2025.139693}{DOI: 10.1016/j.physletb.2025.139693} 

\vspace{2pt}


\end{abstract}

\maketitle

\section{Introduction}

In most cases, low-energy nuclear fission results in the fragmentation of a 
 heavy nucleus into two smaller fragments along with a few neutrons. 
 However, in rare instances, tripartition, characterized by the emission of 
 an additional fragment, may occur. The third fragment is typically an alpha particle, and the likelihood of emitting a heavier third fragment decreases sharply with its mass \cite{Gonnenwein2004}. Genuine (or true) ternary fission, defined as the low-energy fission of a nucleus into three fragments of comparable mass, remains an elusive phenomenon. Although theoretically predicted early on \cite{Present1941}, experimental efforts have since yielded conflicted results. Early studies reported detection rates of approximately $10^{-5}$ per binary fission \cite{Rosen1950, Muga1963, Muga1965}, while others nearly excluded it at levels of $10^{-8}\!\!-\!\!10^{-9}$ \cite{Stoenner1966, Schall1987}. At higher energies (20-100 MeV), evidence of induced ternary fission has been observed \cite{Iyer1968}, and the process becomes more frequent in high-energy collisions (500-600 MeV) \cite{Becker1974, Wilczynski2010}.

 Recent measurements conducted in Dubna \cite{Pyatkov2010, Pyatkov2012, 
 Kamanin2013, Pyatkov2017} suggest that genuine ternary fission may occur
 spontaneously in $^{252}$Cf and be neutron-induced on $^{235}$U. Coincidences
 between two primary fission fragments close to $^{132}$Sn and $^{68,72}$Ni 
 were identified ($4\times 10^{-3}$ of all events), with the missing mass 
 pointing to a potential third fragment such as Ca (or Si in the case of 
 $^{236}$U). The lack of direct detection of the smallest fragment,    
 attributed to its likely low velocity, warrants caution in the 
 interpretation of these results.

Since then, several experimental and theoretical studies have addressed the 
experimental findings of the Dubna group. The process is likely collinear, 
 with the second scission occurring shortly after the first, driven by the 
 large deformation of one primary fragment \cite{Pyatkov2017, Oertzen2020}. 
 Liquid drop model calculations also favor this collinear scenario with respect
 to triangular-configuration fission \cite{Diehl1973, Diehl1974}. Numerous 
 microscopic-macroscopic (MM) studies \cite{Manimaran2011,
 Tashkhodjaev2011, Vijayaraghavan2012, Vijayaraghavan2014, Tashkhodjaev2015,
 Vijayaraghavan2015, Holmvall2017} have aimed to
 shed light on the Dubna results, and Karpov estimated the barrier for
 true ternary fission of $^{252}$Cf to be approximately 7 MeV higher than
 the one for binary fission in a fully consistent MM
 three-center model \cite{Karpov2016}.
 Combined with a random walk at a constant temperature, a MM model yielded 
 a probability of $10^{-5}-10^{-6}$ for the second scission in the spontaneous 
 fission of $^{252}$Cf, but the most probable mass of the small fragment 
 was found to be $A=28$ - smaller than that of Calcium \cite{Wada2021}. 
 
 In this work, we investigate these ternary fission reactions using the 
 Hartree-Fock plus BCS (HFBCS) method with Skyrme-type 
 interactions. We assume, as suggested in previous works, that the shape 
 evolution into final fragments occurs mostly prior to the first scission. 
 The self-consistent approach allows for a better adjustment of the nuclear 
 shape during elongation and neck formation compared to conventional MM 
 methods. We construct energy surfaces for configurations constrained by the 
  required tripartition which reveal a strong force driving the system towards 
 bipartition and limiting the viability of ternary fission paths.  
  By complementing the HFBCS calculations with a 
 simple Langevin-based model and quantum-tunelling probability evaluation, we 
 estimate the relative probability of studied ternary fission reactions with 
 respect to the binary fission. 

\section{The method}

\subsection{General framework}

Our calculations use the SLy6 Skyrme interaction
 \cite{Chabanat1998}, one of the few non-relativistic Skyrme forces which 
 provide a relatively decent description of fission barriers. This is partly due 
 to the inclusion of the two-body contribution to the center-of-mass 
 correction, which effectively improves the surface energy 
 \cite{DaCosta2024}.

The HFBCS equations are solved on a 3D spatial mesh with a step of 0.69 fm, 
scaled to accommodate the system elongation. Our code assumes two plane 
 symmetries (the $xz$ and $yz$ planes), allowing for mass asymmetry exclusively 
 along the $z$-axis. We consider collinear configurations, with the centers of 
 the three fragments aligned along the z-axis. Pairing interactions are 
 incorporated using a delta interaction of fixed strength acting between pairs 
 of time-reversed states. A smooth energy cutoff for the number of interacting 
 levels is applied, following Ref.~\cite{Bender2000}. 
 The BCS solution for 
 state-dependent matrix elements results in pairing gaps that vary across 
 orbitals, typically decreasing with increasing single-particle energy. 
 The pairing strengths are adjusted to reproduce experimental pairing gaps 
 around the Fermi level for $^{252}$Fm, i.e. $\Delta_n=0.696$ MeV and 
 $\Delta_p=0.802$ MeV, setting the corresponding interaction strengths 
 at $V_n=316$ MeV$\cdot$fm$^3$ and $V_p=322$ MeV$\cdot$fm$^3$ respectively. 


The HFBCS equations are solved using imaginary-time evolution, with constraints applied throughout the iterations to enforce specific configurations. 
 In particular, the center of mass of the system is fixed at the 
 origin, and the quadrupole moment $Q_{20}=
 \langle 2z^2-x^2-y^2 \rangle$ is constrained to a prescribed value. The latter is crucial 
 for tracking the elongation of the system and ensuring convergence. 
 The Pauli principle is maintained by keeping the wave functions orthonormal.

\subsection{Tripartition-specific constraints} \label{sec.constraints}

To enforce ternary fission we employ two neck constraints, formulated as a generalization of the approach introduced in Ref. \cite{Berger1990}, specifically: 
\begin{equation}
Q_{n_i}=  \int \! \exp\left[ -\left( \frac{z - z_i}{a_0}\right)^2\right]  \rho({\bf r})\, \dd^3 r , \label{eq.neck}
\end{equation}
where $z_i$, $i=1,2$ are the positions of the necks. Following the prescription
 for binary fission given in Ref. \cite{Han2021}, the diffuseness parameter was
 set at $a_0=1$ fm to ensure a smooth fragmentation while keeping the 
 constraints localized and compatible with the mesh resolution. 

The quantities $Q_{n_i}$ measure the number of nucleons in the vicinity of the 
 neck planes $z=z_i$. Neck formation corresponds to a decrease in $Q_{n_i}$, 
 and genuine tripartition occurs when both $Q_{n_1}$ and $Q_{n_2}$ tend to 
 zero. However, due to the differences in sizes and shapes of the three 
 fragments, equal neck radii may correspond to different constraints 
 $Q_{n_1} \neq Q_{n_2}$. As a result, enforcing $Q_{n_1} = Q_{n_2}$
  is not sufficient to prevent asymmetric necks leading to bipartition. To 
 address this problem, we fix the value $Q^{(0)}_{n_1}$ constrained at the first neck 
  and iteratively determine $Q^{(0)}_{n_2}$ required at the 
 second neck, so that the nucleon counts $N_i=\iiint_{z_i-h/2}^{z_i+h/2}\!\rho({\bf r})\, \dd z \dd x \dd y$, with $h$ the mesh step, become equal. This is done by shifting $Q^{(0)}_{n_2}$ by $N_1-N_2$ at each step. The procedure 
 converges within a few hundred steps, resulting in both necks having similar sizes. 
 After that, the constraint $Q^{(0)}_{n_2}$ is left at the 
 value it reached, i.e. the following HFBCS iterations proceed without further 
 adjustments of $Q^{(0)}_{n_2}$.  
 
An additional constraint $Q_{A_{mid}}=\iiint_{z_1}^{z_2}\!\rho({\bf r})\, \dd z \dd x \dd y $ is applied when the middle fragment loses an excessive number of nucleons, leading the system to bipartition. If the 
 number of nucleons of the middle fragment $Q_{A_{mid}}$ falls below the expected value 
 $A_{mid}$, it is constrained to $A_{mid}$ to avoid binary fission and 
 preserve the intended ternary configuration.

Finally, to facilitate physical interpretation of the results, 
 the maps are generated for fixed neck separations $z_{12} = z_2 - z_1$, which 
 effectively acts as an implicit fourth constraint.

\subsection{"Forbidden" regions} \label{ssec.forbiddenRegions}

\begin{figure}[htb]
  \center
  \includegraphics[width=0.95\columnwidth]{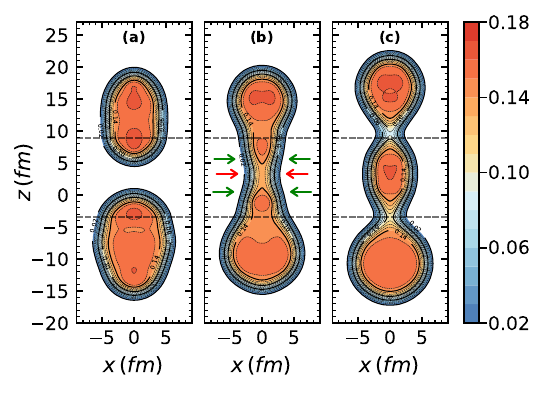}
  \caption{Density distributions of the $^{252}$Cf system for the neck distance $d_{^{48}\mbox{\scriptsize Ca}} + 4$fm (see Sec. \ref{ssec.PES}) The neck positions are indicated by dashed lines. Panel (a) shows a forbidden bipartite shape, 
 which still satisfies both neck constraints and the nucleon number 
 of the middle fragment. Panel (b) illustrates the density criterion: 
 an axial density $\rho(0, 0,z)$ on the z-axis below 0.14 fm$^{-3}$ 
 indicates a tendency towards bipartition (red arrow) . The convex shape is recovered where the axial density becomes larger (green arrows). Panel (c) shows the
  configuration at the peak of the tripartition barrier.}
  \label{fig.densities}
\end{figure}

  The constraints defined above do not always enforce tripartition.
  There are cases in which, in spite of the constraints, the resulting 
 densities show bipartition or a clear tendency towards it.  Fig. 
 \ref{fig.densities}(a) illustrates such an example of a matter density which 
 satisfies the imposed conditions - namely the appropriate nucleon count in 
 the middle fragment and correctly constrained neck nucleon numbers, the neck 
 positions being represented as dashed lines. Yet, rather that evolving as a 
 single elongated nucleus towards ternary fission, the system separates into 
 two fragments which subsequently reposition along the $z$-axis to meet 
 the constraints. 

To identify configurations tending towards bipartition, 
 which we will refer to as "forbidden", we introduce two criteria that help interpreting energy maps and 
  tracing plausible ternary fission paths. While 
 subjective to a degree, they enable automatic classification of the
 HFBCS solutions, and will be later validated by 
 probability estimates. 
 
 The first condition applies when the system begins to develop 
  a neck between the designated $z=z_i$ neck planes 
 (see Fig. \ref{fig.densities}(b)). We noticed that this associates with a
  nuclear density on the $z$ axis dropping below the typical saturation density
  of 0.15$-$0.16~fm$^{-3}$. Hence we identify emerging bipartition if 
   $\rho(0,0,z)<0.14$ fm$^{-3}$ within the region $z_1<z<z_2$.  
 As illustrated in Fig.  \ref{fig.densities}(b), the densities below this value
 are associated with a local concavity of the nuclear surface (red arrow), 
  whereas higher densities preserve a convex surface (green arrows). 

 
 The second condition pertains to highly elongated systems in which
 one of the two necks has ruptured. Similarly to the case shown in Fig.  
 \ref{fig.densities}(a), once a neck breaks, the remaining neck constraint no 
 longer corresponds to a true "neck", as the detached fragments simply shift 
 up and down along the $z-$axis to fulfill it. In practice, two necks are never
  identical, and one always breaks first. To avoid misinterpreting such 
 asymmetries, we impose the following criterion: if only one of the two necks 
 is ruptured across 
 two consecutive mesh points $(Q_{20},Q_{n_1})$, $(Q_{20},Q_{n_1}')$ with 
 $Q_{n_1}'<Q_{n_1}$, we classify the intermediate region as leading to 
 bipartition. While sequential tripartition could still occur depending on the 
 thickness of the remaining neck, such scenarios lie beyond the scope of our 
 analysis based on configurations of the initial heavy nucleus. We consider 
 a density $\rho(0,0,z)< 0.08$ fm$^{-3}$ (half the saturation density) anywhere
  on the $z$-axis as indicative of scission.

\subsection{Dynamical suppression of ternary fission} \label{ssec.probabilities}

As our HFBCS results suggest, the ternary fissions of interest occur at very 
 large elongations ($Q_{20}\approx 650-750$ b) and remain in persistent 
 competition with bipartition along the trajectory.  
 A static description of their probability based solely on the quantum 
 tunneling of the energy barrier thus fails to account for the dynamical suppression 
 along the path.

To estimate this suppression, we use a Langevin-based model, 
 in which the probability of a given trajectory is explicitly determined by the proper action integral \cite{Onsager1953, Fitzgerald2023}. Given a set of reduced (dimensionless) collective variables 
 $\bm{X}=\left\{X_k\right\}$, we assume the overdamped dynamics defined by 
  the equations of motion: 
 \begin{equation}
\label{eq.Langevin}
   {\bm {\dot X}} = -\frac{1}{\Gamma}\nabla_{\bm X} V + {\vec \xi}(t) ,
\end{equation}
where $V(\bm{X})$ is the potential energy, $\Gamma$ is the damping coefficient.
 The random force ${\vec \xi(t)}$ has zero mean and satisfies the correlation 
 relation $\langle\xi_i(t)\xi_j(t')\rangle=2D\delta_{i j}\delta(t-t')$ with 
 $D=T/\Gamma$ and $T$ is the temperature in MeV. 

 The temperature of an excited nucleus is usually assumed to be
 $\tilde{T}=\sqrt{E^*/a}$, where $E^*$ is the internal excitation energy,
  $a=A/n $ MeV$^{-1}$ is the level density parameter with $n = 8 - 12$.
 Along tripartition paths, one has $E^* \lesssim 10-$20 MeV, yielding a 
 temperature $\tilde{T}\lesssim 0.6-$0.9 MeV that varies along the trajectory. 
 Here, we make use of the quantum-corrected temperature \cite{Hofmann1977, 
 Hasse1979}, which remains nearly constant for the present excitation energies 
 and corresponds to the zero-point collective energy of the 
  heat-bath oscillators, i.e. $T \simeq E_0$, with $E_0 = 1.5-2$ MeV for 
  heavy and superheavy nuclei \cite{Pomorski2023, Jaganathen2025}. 
 This value higher than ${\tilde T}$ thus provides an upper bound for 
 probability. Moreover, for simplicity, we assume a constant damping 
  $\Gamma$. 
 To maintain internal consistency with this assumption, the variables $X_i$ 
 must be commensurable.  
 We thus transform the constrained variables to lengths and 
 define the reduced collective variables $X_i$ as their ratios to 
  characteristic reference lengths. Specifically, we define: 
\begin{align}
L_{20} & = \frac{2}{\sqrt{3}} \sqrt{\frac{Q_{20}}{A_{tot}} + <r^2>} = 
 2\sqrt{<z^2>}, \\
L_n & = \frac{2}{\pi^{3/4}}\sqrt{\frac{Q_{n_1}}{a_0\rho_0}} , \label{eq.langevin}
\end{align}
where $<r^2>$ is the calculated rms radius of the system, $A_{tot}$ the total 
 number of 
 nucleons and $\rho_0=0.16$ fm$^{-3}$ is the nucleonic density. $L_{20}$ 
 represents the total system length along the z-axis, and $L_n$ an approximate 
 diameter of the neck. If the middle fragment is approximated as a cylinder 
 with the height $L_z=z_{12}$, its effective diameter $L_{mid}$ is related to 
 $A_{mid}$ via $L_{mid}=2\sqrt{A_{mid}/(\pi \rho_0 z_{12})}$.  
 Since $L_{20}$ and $L_n$ characterize the whole system and $L_{mid}$ and 
 $L_{z}$ are specific to the middle fragment, we define: 
\begin{align}
X_{20} & = L_{20} / d_{A_{tot}} , &X_{n} & = L_{n} / d_{A_{tot}}, \\
X_{mid} & = L_{mid} / d_{A_{mid}} , &X_{z} & = L_{z} / d_{A_{mid}},
\end{align}
where $d_A = 2r_0A^{1/3}$ is the diameter of the corresponding spherical 
 nucleus and $r_0 = 1.16$ fm. 
   
Following Refs. \cite{Onsager1953, 
 Fitzgerald2023}, the probability of a trajectory ${\bm X}(t)$
 developing under Eq. (\ref{eq.Langevin}) from  
 $\bm{X}(t_i)=\bm{X}_i$ to $\bm{X}(t_f)=\bm{X}_f$ over the time interval 
 $\Delta t = t_f - t_i$ is proportional to: 
\begin{equation}
P(\bm{X}_i \!\rightarrow\! \bm{X}_f) \propto \exp \!\left[ -\frac{1}{4\Gamma T} 
 \!\!\int_{t_i}^{t_f} \!\! \left(\Gamma\dot{\bm{X}} + \nabla_{\!\!\bm{X}}V 
 \right) ^2\! \dd t\right], \label{eq.finalProbability0}
\end{equation}
where $\bm{X}^2 := \bm{X}^T \bm{X} $. 
 For a fixed path, the action is minimal for $|\dot{\bm X}|=
 |\nabla_{\bm X} V|/\Gamma$ which gives the maximal probability: 
\begin{equation}
P(\bm{X}_i \!\rightarrow\! \bm{X}_f) \propto \exp \!\left[ -\frac{1}{2T}\left( \Delta V +  \!\!\int_{\bm{X}_i}^{\bm{X}_f} \vert \nabla_{\bm{X}} V\vert \, \dd s
 \right)\right], \label{eq.finalProbability1}
\end{equation}
 with $\dd s=|{\dot {\bm X}}| \dd t$, $\Delta V = V(\bm{X}_f) - 
 V(\bm{X}_i)$. With a constant damping coefficient, this 
 probability is independent of friction itself, which is 
 instead absorbed into the traversal time: 
\begin{equation}
\Delta t_{tr} = \Gamma \int_{\bm{X}_i}^{\bm{X}_f} \! \frac{1}
 {\vert \nabla_{\bm{X}} V\vert} \dd s. \label{eq.time}
\end{equation}

The probabilities in Eqs. (\ref{eq.finalProbability0}, 
 \ref{eq.finalProbability1}) are only determined up to a prefactor, which 
 would typically be obtained by summing over all possible trajectories and 
 normalizing the total probability to one. However, this prefactor approximately cancels out when
 considering relative probabilities. 
Assuming that nearly all fission paths originating behind the standard 
 fission barrier lead to bipartition, with tripartition forming only a 
 minuscule fraction, we can estimate the relative 
 probability of tripartition with respect to bipartition along a single, most 
 probable trajectory selected from a few tested candidates.


\section{Results and discussions} \label{sec.results}

\subsection{Potential energy surfaces of $^{252}$Cf and $^{236}$U} \label{ssec.PES}

\begin{figure*}[ht]
  \centering
  \includegraphics[width=1.0\textwidth]{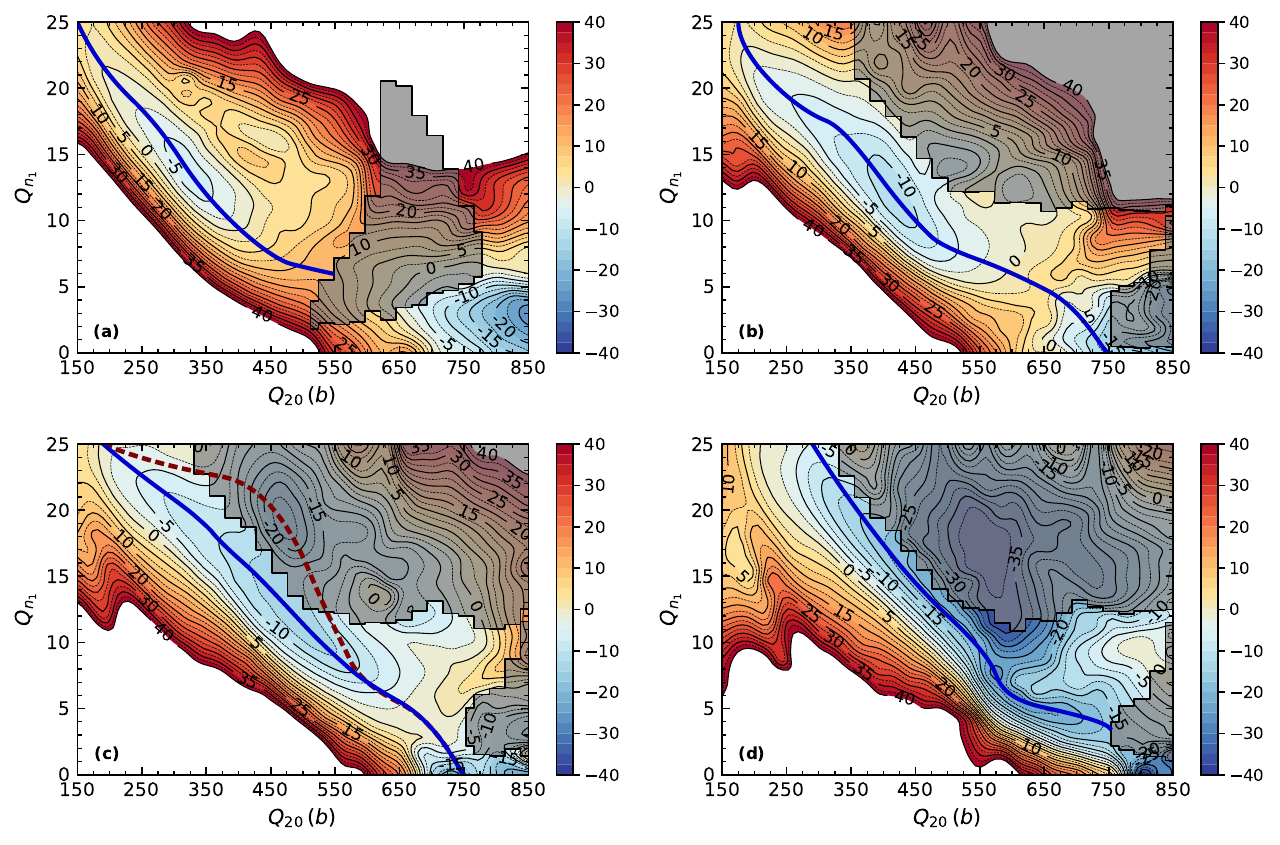}
  \caption{Potential energy surfaces with respect to the ground-state energy 
 for the tripartition reaction $^{252}$Cf $\rightarrow  ^{132}$Sn + $^{48}$Ca +
 $^{72}$Ni at the neck distances $d+ \delta$, $\delta = 0, 2, 4, 6$ fm in panels 
 (a)$-$(d), respectively. The shaded areas indicate forbidden regions (see text). 
 The plausible trajectories are shown in blue.}
  \label{fig.maps252Cf}
\end{figure*}

\begin{figure*}[ht]
  \centering
  \includegraphics[width=1.0\textwidth]{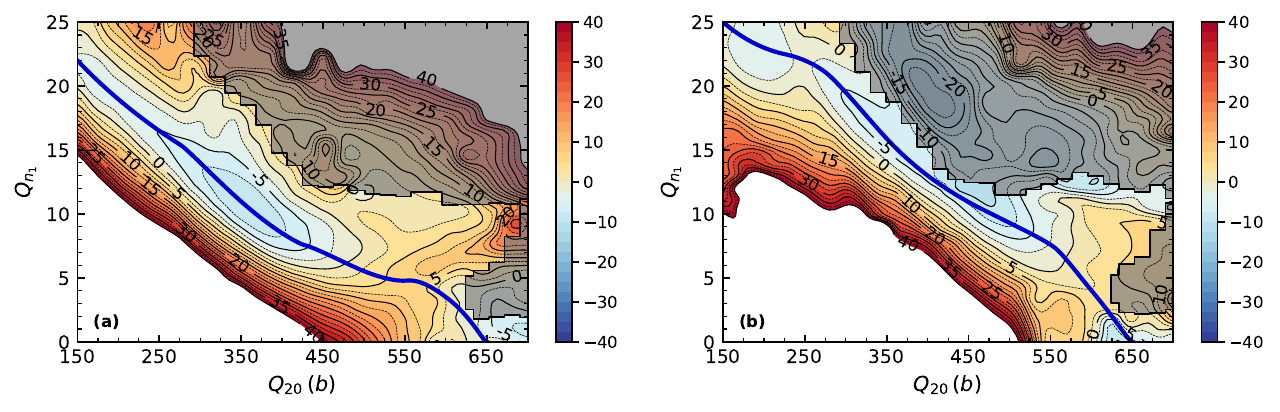}
  \caption{Same as Fig. \ref{fig.maps252Cf}, but for the reaction 
 $^{236}$U $\rightarrow  ^{132}$Sn + $^{34}$Si + $^{70}$Ni at the neck 
 distances $d + 2$ fm (a) and $d + 4$ fm (b). The energies are shown with 
 respect by the energy of the first (binary) fission barrier, corresponding to 
 the neutron-induced fission.}
  \label{fig.maps236U}
\end{figure*}

The framework was applied to the reactions: 
\begin{align}
\prescript{252}{98}{\mbox{Cf}}_{154} & \rightarrow \prescript{132}{50}{\mbox{Sn}_{82}} + \prescript{48}{20}{\mbox{Ca}}_{28} + \prescript{72}{28}{\mbox{Ni}}_{44} ,\\
 \prescript{236}{92}{\mbox{U}}_{144} & \rightarrow \prescript{132}{50}{\mbox{Sn}}_{82} + \prescript{34}{14}{\mbox{Si}}_{20} + \prescript{70}{28}{\mbox{Ni}}_{42} .
\end{align}
In the case of $^{252}$Cf, five of the six nucleon numbers of fragments 
 are magic (20, 28, 50, 82), which makes the subsystems highly stable and 
 might favor tripartition. The potential energy surfaces (PES) in 
 Figs. (\ref{fig.maps252Cf}, \ref{fig.maps236U}) were computed on a 
 $30 \times 30$ mesh with $Q_{20} = 150-850$ b for $^{252}$Cf, 
 $Q_{20} = 150-700$ b for $^{236}$U, and $Q_{{n_1}} = 0-25$ in both cases. 
 For reference, standard configurations of $^{252}$Cf and $^{236}$U at the 
 elongation $Q_{20}=150$ b corresponds to $Q_{n_1} \approx 25$. Tripartition 
 occurs in the lower-right parts of the maps, at small neck values $Q_{{n}} 
 \lesssim 2-3 $ and large elongations. For readability, the PES are 
 shifted with respect to a reference energy $E_{\mbox{\footnotesize ref}}$, defined as the 
 ground-state energy for $^{252}$Cf or the energy of the first barrier for 
 $^{236}$U, corresponding to the neutron-induced fission.

For $^{252}$Cf, the PES were computed for four fixed neck distances 
 $z_{12} = d_{^{48}\mbox{\footnotesize Ca}} + \delta$ with $\delta = 0, 2, 4, 6$ fm and $d_{^{48}\mbox{\footnotesize Ca}} = 8.43$ fm calculated with $r_0=1.16$ fm. For reference, within HFBCS, the rms diameter of the 
 spherical $^{48}$Ca is 9.09 fm, i.e. $d_{^{48}\mbox{\footnotesize Ca}} 
 + 0.66$ fm. The neck positions are fixed by the geometry of three spheres of 
 diameters $d_{A_1}$, $d_{^{48}\mbox{\footnotesize Ca}}$ and $d_{A_3}$ ($A_1 
 = 132$ and $A_3=72$), in a linear touching configuration, with the center of 
 mass positioned at the origin.


The final constraint values align with the imposed ones to within 0.01\% 
 for $Q_{20}$, and within $0.5-1$ nucleon for $Q_{n_i}$ and 
 $Q_{A_{mid}}$. We then interpolate the values back onto the mesh points using 
 a barycentric (linear) interpolation over triangular cells containing the mesh 
 points, which is accurate given the close agreement with the target 
 constraints. 
 
Shaded regions in Fig. \ref{fig.maps252Cf} indicate the forbidden 
 configurations 
 meeting the criteria defined in Sec. \ref{ssec.forbiddenRegions}: the upper 
 areas correspond to the formation of a neck in between $z_1$ and $z_2$ despite the tripartition 
 constraints, while the lower areas reflect the premature rupture of 
 a single neck.
Plausible ternary fission trajectories that avoid these forbidden regions
 and follow the energy valley in $(Q_{20}, Q_{n_1})$ as closely as possible
  are shown in blue.
  
 At the smallest neck distance $z_{12} = d + 0$ fm, the middle fragment is 
 too compact for a neck to form between $z_1$ and $z_2$, hence the absence 
 of a top shaded area. Instead, the system favors an 
 early rupture of one neck, usually leaving one nucleus close to  
 $^{132}$Sn and a moderately deformed $A\approx 120$ fragment, making 
 genuine tripartition improbable. Conversely, at 
 $\delta = 6$ fm (panel (d)), along 
 tripartition trajectories, the system is pushed by the energy gradients 
 in $Q_{20}$ and $Q_{n_1}$ towards the upper shaded area representing bipartition: the 
 very elongated middle fragment easily breaks in the middle. Thus, tripartition, if 
 it occurs at all, seems most likely at intermediate neck distances.
 The path shown for $\delta=2$ fm (panel (b)) follows what appears to be a 
 ternary fission valley, whereas a noticeable force towards the bipartition 
 region emerges at $\delta=4$ fm. It is important to note that the 
 PES are shaped by a strong constraint on $A_{mid}$. The observed proximity 
 between the tripartition valley and the top shaded bipartition area thus 
 indicates that the formation of a central $^{48}$Ca fragment is dynamically 
 suppressed and happens only at very large system elongations.

\begin{figure}[htb]
  \center
  \includegraphics[width=0.95\columnwidth]{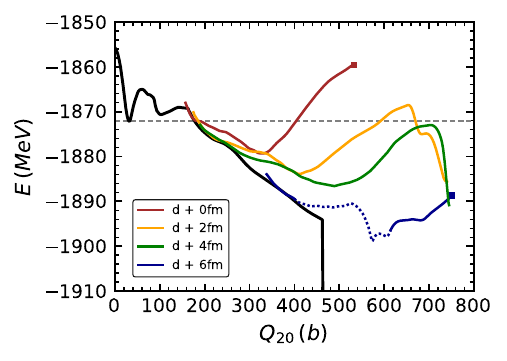}
  \caption{Energies along the trajectories from Fig. \ref{fig.maps252Cf}
 for the reaction $^{252}$Cf $\rightarrow ^{132}$Sn + $^{48}$Ca + $^{72}$Ni 
 at various neck distances. The horizontal dashed line marks the g.s. energy. 
 The dashed trajectory segments indicate a strong force towards bipartition 
 and square markers mark a premature rupture of a single neck. The 
 super-elongated (binary) fission path is shown in black.}
  \label{fig.energies252Cf}
\end{figure}

Fig. \ref{fig.energies252Cf} shows the energy profiles along the trajectories 
 from Fig. \ref{fig.maps252Cf}, as well as along the super-elongated 
 path obtained from standard HFBCS calculations using only the $Q_{20}$
 constraint, but initiated from configurations with two pronounced necks. 
 This super-elongated path, reaching bipartite fission at $Q_{20} 
 \approx 460$ b, differs from the conventional bipartition path that leads to 
 scission at $Q_{20} \approx 250$ b. Remarkably, the tripartition 
 trajectories connect to this path.

 The energies along the ternary paths decrease with the elongation of the 
 middle fragment (increasing $\delta$). For $\delta = 4$ fm, ternary fission 
 can proceed without tunneling, as the barrier lies below the ground state. 
 However, as discussed in the next section, it remains dynamically hindered.

\begin{figure}[htb]
  \center
  \includegraphics[width=0.95\columnwidth]{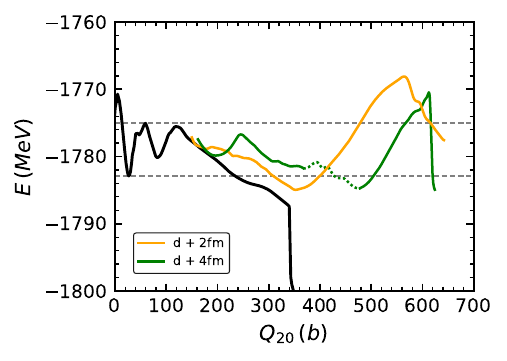}
  \caption{Same as Fig. \ref{fig.energies252Cf}, but for the reaction 
 $^{236}$U $\rightarrow \: ^{132}$Sn + $^{34}$Si + $^{70}$Ni. Here, the 
 horizontal dashed lines mark the energies of the g.s. and the (binary) fission
  barrier $E_{\mbox{\footnotesize ref}}$.}
  \label{fig.energies236U}
\end{figure}

We repeated the analysis for the $^{236}$U system using the neck distances 
 $d_{^{34}\mbox{\footnotesize Si}}+ \delta$, $\delta=2,4$ fm (Figs. \ref{fig.maps236U}, 
 \ref{fig.energies236U}).

 While the overall trends are similar, the plausible ternary fission paths 
 exhibit energy barriers of 4.4 MeV (green) and 7.0 MeV (yellow) for 
 $\delta = 4$ fm and $\delta = 2$ fm, respectively. Interestingly, parts of the
 $\delta = 4$ fm ternary path lie above the $\delta = 2$ fm path in energy, 
 suggesting that the formation of the middle $^{34}$Si fragment may involve 
 a gradual increase in neck distance.
 

\subsection{Estimate of tripartition probabilities}

 To quantify the dynamical suppression of true ternary fission, the PES were first transformed to $(L_{20},L_n)$ coordinates and 
 the energy gradients with respect to $L_{20}$ and $L_{n}$ were obtained by 
 cubic B-spline interpolation over the PES meshes. For the neck distance 
 $z_{12}$ and the middle fragment mass $A_{mid}$, finite-difference  
 approximations were used with additional PES maps computed respectively at 
 $z_{12} \pm 2$ fm and $A_{mid} \pm 4$. While the constraint on $A_{mid}$ is 
 only active when it falls below its target value, we observe that it switches 
 on and off abruptly. In regions where the constraint is inactive, the related 
 gradient is zero. The partial derivatives with respect to $z_{12}$ and 
 $A_{mid}$ obtained from finite differences at each mesh point were then 
 interpolated by cubic B-splines along the trajectories. 
 All gradients were subsequently expressed in terms of the reduced variables 
 $X_i$.

The tripartition probabilities per binary fission were calculated for the neck 
 distances $d$ + 2 and $d$ + 4 fm along the trajectories drawn on Figs. 
 \ref{fig.maps252Cf} and \ref{fig.maps236U}. 
 The integrals in Eq. (\ref{eq.finalProbability1}) were evaluated between an 
 initial point ${\bm{X}}_i$ seen on the maps, and a final point ${\bm{X}}_f$ 
 defined by the condition $V({\bm{X}}_f)=E_{\mbox{\footnotesize ref}}$ when 
 an energy barrier exists above $E_{\mbox{\footnotesize ref}}$. Otherwise, the 
 integrals extend to a point just beyond the barrier peak. We thus exclude 
 sub-barrier segments of fission paths from the integral 
  (\ref{eq.finalProbability1}), making  $\Delta V$ either negligible or 
 zero. In most estimates, we used a temperature $T=2$ MeV, and the damping 
 coefficient $\Gamma$ = 100 $\hbar$ to 
 compute the tripartition timescale $\Delta t_{tr}$ (see Ref. 
 \cite{Jaganathen2025} for a typical range of $\Gamma$ values).

 For $^{252}$Cf at $z_{12}=d+4$ fm, we estimate the relative ternary fission 
 probability $P = 3.4 \times 10^{-8}$, with a time $\Delta t_{tr} \sim 8.9 \times 
 10^{-22}$ s within expected fission time scales. 
 For comparison, at $T=1.5$ MeV  
 our estimate drops to $P=1.1 \times 10^{-10}$.  
 These values illustrate  the strong dynamical suppression of tripartition even
 in the absence of a potential barrier. This suppression is mainly tied
 to the $Q_{A_{mid}}$ constraint that enforces the formation of middle fragment:
 omitting the corresponding gradient indeed increases the probability 
 to $P = 4.7 \times 10^{-5}$ (at $T=$ 2 MeV). 
 For a path leading through the forbidden region (maroon dashed 
 trajectory in Fig.~\ref{fig.maps252Cf}(c)), the probability falls to 
 $P = 4.8 \times 10^{-16}$, 
 thereby validating our criteria for the shaded areas.
 
In the remaining cases, quantum tunneling through the energy barrier must also 
 be taken into account in the probability estimate. To this end, we 
 intentionally overestimate the transmission probability $P_t$, while 
 simplifying the dynamics: we only consider the change in the distance between 
 two primary fragments, approximated by $\Delta L_{20}$, and use the 
 reduced mass $\bar{\mu} M_n$ as the (smallest possible) mass parameter of the 
 collective motion, with $\bar{\mu}=A_1(A_{mid}+A_3)/A_{tot}$ or 
 $A_3(A_1+A_{mid})/A_{tot}$, and $M_n$ the average nucleon mass. 
 The tunneling probability is approximated by $P_t\approx \exp(-2S/\hbar)$, 
 where the action is $S=\sqrt{2\bar{\mu} M_n}\int \!
 \dd L_{20}\sqrt{V-E_{\mbox{\footnotesize ref}}}$.  
 
 For $^{252}$Cf at $z_{12} = d + 2$ fm, our calculations yield a dynamical 
 suppression of $P_d = 2.1 \times 10^{-6}$ (with $\Delta t_{tr}=7.4 \times 
 10^{-22}$ s), combined with an additional quantum tunneling suppression of 
 $P_t \approx 10^{-3}$ through the $3.5$ MeV energy barrier observed in 
 Fig. \ref{fig.energies252Cf}.
 
 In the case of $^{236}$U, for the neck distance $z_{12}=d+4$ fm, the 
 calculated dynamical probability is $P_d=1.7 \times 10^{-8}$, 
 further suppressed by $P_t \approx 10^{-2}$ due to the tunneling of the 
 remaining 4.4 MeV barrier (see Fig. \ref{fig.energies236U}). 
  For $z_{12}=d+2$ fm, $P_d = 2.0\times 10^{-5}$, but the remaining 7.0 MeV 
 barrier leads to a strong quantum suppression of $P_t\approx 10^{-7}$.

\section{Conclusions}

 We have computed HFBCS potential energy surfaces for   
 the ternary decays $^{252}$Cf $\rightarrow ^{132}$Sn + $^{48}$Ca + $^{72}$Ni 
 (spontaneous) and $^{236}$U$\rightarrow ^{132}$Sn + $^{34}$Si + 
 $^{70}$Ni (neutron-induced). The imposed constraints do not prevent configurations 
 tending towards bipartition to occupy substantial regions 
 of maps, and energy gradients tend to push plausible ternary fission paths towards them. At intermediate elongations $z_{12}$ of the middle fragment, a ternary fission valley 
 appears, held by a strong constraint on the mass of the middle fragment. 
  As $z_{12}$ increases, the energy barriers against ternary fission diminish, such 
 that tripartition in $^{252}$Cf may proceed without barrier tunneling for sufficiently large $z_{12}$.     
 However, our estimates based on the simple Langevin-type model and 
 simplified account for quantum tunneling 
  show that the studied reactions are either dynamically or quantally suppressed.   
 Despite assumptions overestimating the probability of the sought ternary fission, we obtain probabilities relative to binary fission as low as 
 $10^{-8}-10^{-9}$ for $^{252}$Cf and even less, $10^{-10}-10^{-11}$ for 
 $^{236}$U.
 
 These results emphasize the importance of a dynamical description,
  which does not assume quantum tunneling over a barrier as the only 
 impediment to ternary fission. The framework presented here provides a 
 foundation for future studies of tripartition in superheavy nuclei, where 
 static conditions are much more favorable but for which the dynamical suppression 
 might also remain a critical obstacle.

\section*{ACKNOWLEDGMENTS}

We extend our gratitude to the CIŚ computing center at the National Centre for Nuclear Research for providing extensive computation resources essential to this study. 

\bibliographystyle{model1a-num-names.bst}
\bibliography{Tripartition_bibfile}

\end{document}